\documentclass[aps,pre,preprint,groupedaddress,showpacs]{revtex4}
\usepackage{graphicx}
\begin{document}
\title{Quasi-static relaxation of arbitrarily shaped sessile drops}
\author{Stanimir Iliev}
\email[email:]{ stani@imbm.bas.bg}
\author{Nina Pesheva}
\affiliation{Institute of Mechanics,
Bulgarian Academy of Sciences, \\
Acad. G. Bonchev St. 4, 1113 Sofia, Bulgaria}
\author{Vadim S.Nikolayev}
\affiliation{ESEME, Service des Basses Temp\'eratures,
CEA-Grenoble, France}
\altaffiliation[Mailing address:\ ] {CEA-ESEME, PMMH,
 ESPCI, 10, rue Vauquelin, 75231 Paris Cedex
5, France}
\date{\today}
\begin{abstract}
We study a spontaneous relaxation dynamics of arbitrarily shaped
liquid drops on solid surfaces in the partial wetting regime. It
is assumed that the energy dissipated near the contact line is
much larger than that in the bulk of the fluid. We have shown
rigorously in the case of quasi-static relaxation using the
standard mechanical description of dissipative system dynamics
that the  introduction of a  dissipation term proportional to the
contact line length leads to the well known local relation between
the contact line velocity and the dynamic contact angle at every
point of an arbitrary contact line shape. A numerical code is
developed for 3D drops to study the dependence of the relaxation
dynamics on the initial drop shape. The available asymptotic
solutions are tested against the obtained numerical data. We show
how the relaxation at a given point of the contact line is
influenced by the dynamics of the whole drop which is a
manifestation of the non-local character of the contact line
relaxation.
\end{abstract}
\maketitle

\section{Introduction}

The spreading of a liquid drop deposited on a solid substrate has
many technological applications stimulating active research on
acquiring accurate knowledge on its relaxation that follows the
deposition. More specifically, one is interested to know how the
relaxation rate depends on the initial drop shape and the
properties of the contacting media. It is a complex theoretical
problem and there are numerous studies devoted to drop relaxation
using different approaches and techniques: e.g. macroscopic
\cite{Huh,V76,D79,T79,deG,HM,Cox86} and more recent microscopic
approaches using molecular dynamic simulations
\cite{BCCR97},\cite{RBC99} and Monte-Carlo simulations of 3D
lattice gas \cite{Tan} and 2D \cite{CE93} and 3D Ising model
\cite{NJ04} etc., to mention just few of them.

In the case of  partial wetting  this problem turns out to be very
difficult because of the presence of the triple gas-liquid-solid
contact line. Since the work \cite{Huh}, it has become obvious
that the contact line motion cannot be described with the
classical viscous hydrodynamics approach that uses the no-slip
boundary condition at the solid surface. The velocity ambiguity at
the moving contact line leads to the un-physical divergences of
the hydrodynamic pressure and viscous dissipation. Multiple
approaches were suggested to overcome this problem. Among the most
popular solutions one can name a geometrical cut-off \cite{deG} or
the local introduction of the slip near the contact line
\cite{HM}. One finds experimentally \cite{JFM02,NBN} that while
the dissipation is finite, it is very large with respect to the
bulk viscous dissipation. Several physical mechanisms are
suggested to describe the contact line motion \cite{BH1,YS}.

Following a suggestion in \cite{deG}, a combined approach was
proposed in \cite{RCO99} considering both, the bulk viscous
dissipation and the dissipation occurring at the moving contact
line, to study the drop relaxation in the partial wetting regime.
A phenomenological dissipation per unit contact line length was
introduced. It was taken to be proportional to the square of the
contact line velocity $v_n$ (the first term symmetric in $v_n$) in
the direction normal to the contact line. There the  standard
mechanical description of dissipative system dynamics  was applied
to describe  the time evolution of the drop contact line in the
case of a spherical cap approximation for the drop shape in the
quasi-static regime. Considering the drop as a purely mechanical
system, the driving force for the drop spreading was balanced by
the rate of total dissipation. No assumption was made for a
particular line motion mechanism.

This approach was further generalized to any contact line
shape in \cite{VNDB1}
by writing  the energy dissipated in the system per unit time as
\begin{eqnarray}
T=\oint{\xi v_n^2 \over 2}\, dl  \label{dfv} \ ,
\end{eqnarray}
where the integration is over the contact line of the drop and $\xi$
is the dissipation coefficient.

In the present work we employ this approach to study the
quasi-static relaxation of arbitrarily shaped drops in the partial
wetting regime. It is assumed here that the energy dissipated near
the contact line is much larger than that in the bulk of the
fluid. In Section II we show that this approach actually leads to
the local relation (first obtained in the molecular-kinetic model
of contact line motion of Blake and Haynes \cite{BH1}) between the
contact line velocity and the dynamic contact angle at every point
of an arbitrarily shaped contact line. In Section III we describe
a numerical 3D code for studying the relaxation of an arbitrarily
shaped drops starting directly from the variational principle of
Hamilton, taking into account the friction dissipation during the
contact line motion. In Section IV we obtain numerically and
discuss the quasi-static relaxation of a drop with different
initial shapes. Section V deals with our conclusions.

\section{THE MODEL}

We consider a model system consisting of  a 3D liquid drop placed on a
horizontal, flat and chemically homogeneous solid substrate.
Both the drop and the substrate are surrounded by an
ambient gas and it is assumed that the liquid and the gas are mutually
immiscible. Initially the drop deposited on the substrate is out
of equilibrium. Under the action of the surface tension, the
incompressible liquid drop relaxes towards spherical cap shape.
The drop is assumed to be small enough so that the influence of
the gravitation on its shape can be neglected. According to the
capillary theory \cite{F86,LL87}, the potential energy of the system is:
\begin{eqnarray}
   U =A_{lg} \sigma_{lg}+A_{ls} \sigma_{ls}+A_{sg} \sigma_{sg} ,
\label{en}
\end{eqnarray}
where  the  surfaces $A_{lg}$, $A_{ls}$, and $A_{sg}$ (with
corresponding surface tensions $\sigma_{lg}$, $\sigma_{ls}$, and
$\sigma_{sg}$) separate the liquid/gas, liquid/solid, and
solid/gas phases respectively. In accordance with the approach
described in \cite{RCO99,BH1,VNDB1}, we assume that with the
moving contact line a dissipation function $T$ is related, given
by Eq.(\ref{dfv}).

According to the variational principle of Hamilton one writes:
 \begin{eqnarray}
 \int_{t_0}^{t_1}(\delta K+\delta W ) \, dt \ =0 ,
 \label{var}
 \end{eqnarray}
where $\delta W$ is the virtual work of the active forces and
$\delta K$ is the variation of the kinetic energy of the system.
The virtual work is $\delta W =-\delta U+\delta W_1 $, where
$\delta W_1$ is the virtual work related  to the friction
dissipation  Eq.~(\ref{dfv}). A class of virtual displacements is
considered in Eq.~(\ref{var}) satisfying the  conditions of
immiscibility, of conservation of the area of the solid surface
and the condition of constant volume $V$. Since the Lagrangian  is
${\mathcal L}=K-U$, the variational condition given by
Eq.~(\ref{var}) can be put in the following form:
 \begin{eqnarray}
\int_{t_0}^{t_1} (\delta  {\mathcal L}+ \delta  W_1)\, dt \ =0 .
 \label{var2}
 \end{eqnarray}
The contribution of the kinetic energy of the fluid motion is
assumed to be negligible because we consider a quasi-static
relaxation here, so that $\mathcal L=-U$.

The radius-vectors $\vec R$ of the points of the liquid/gas
interface $A_{lg}$ are taken as generalized coordinates. These
coordinates are not independent, their displacements have to
satisfy the condition of constant drop volume. Taking into account
this condition by introducing a Lagrange multiplier $\lambda$ and
adding the term $\lambda V$ into Eq.~(\ref{var2}) one obtains:
\begin{eqnarray}
\int_{t_0}^{t_1}(-\delta U+\delta W_1+\lambda \delta V) \, dt \ =0 .
\label{var3}
\end{eqnarray}
 The Lagrange multiplier $\lambda$ (its physical meaning is the pressure
 jump across the drop surface $A_{lg}$) varies in time.
 So in the quasi-static regime one has the following equation
 \begin{eqnarray}
-\delta U+\delta W_1+\lambda \delta V \ =0 , \label{var3p}
\end{eqnarray}
 where $\delta W_1$ is given by (see, e.g., \cite{LL69})
\begin{eqnarray}
 \delta W_1=-\xi \oint_L v_n \delta {\vec R}\,dl\, .
 \label{vl2s}
 \end{eqnarray}
The variation of the potential energy under the constant volume
constraint reads \cite{F86}
 \begin{eqnarray}
 \delta  (-{U}+\lambda V)=\int_{A_{lg}}
( 2 \sigma_{lg}k-\lambda) \delta {\vec R} \,
dA_{lg}+ \sigma_{lg}\oint_L (\cos \theta_{eq} -\cos
\theta)\delta {\vec R}\, dl,
 \label{vl2}
 \end{eqnarray}
where $k$ is the mean curvature of the liquid/gas interface $A_{lg}$;
 $\delta\vec R$ is the virtual displacement
of the points normal to the drop surface $A_{lg}$ in the first and
to the contact line $L$ in the second integrals respectively;
$\theta$ is the dynamic contact angle, $\theta_{eq}$ is the
equilibrium contact angle defined by the well known Young
equation:
 \begin{eqnarray}
  \cos \theta_{eq}=(\sigma_{gs}-\sigma_{ls})/\sigma_{lg}.
  \label{young}
 \end{eqnarray}
Substituting Eqs.~(\ref{vl2s}) and (\ref{vl2}) in
Eq.~(\ref{var3p}) and taking into account the independence of the
virtual displacements of the points of the interface $A_{lg}$ and
of the contact line $L$ (due to which each of the integrands must
be equated to zero separately), one obtains
 the Laplace equation
  \begin{equation}\label{Lap}
 -2 \sigma_{lg}k+\lambda=0,
\end{equation}
from which the surface shape can be obtained at any time moment and the
equation
 \begin{eqnarray}
 (\cos \theta_{eq}-\cos \theta({\vec R}))=
{\xi \over  \sigma_{lg}}
 {\vec v}_n ({\vec R})\, ,
 \ {\vec R} \in L \
 \label{clv}
 \end{eqnarray}
valid at the contact line. Eq.~(\ref{clv}) serves as a boundary
condition for Eq.~(\ref{Lap}). For a given volume $V$ and
arbitrary initial contact line position $L_0$,
Eqs.~(\ref{Lap},\ref{clv}) define the evolution of the drop shape
and of the drop contact line. However, in our calculations we will
not use Eqs.~(\ref{Lap},\ref{clv}) directly, we will use
Eqs.~(\ref{var3p}, \ref{vl2s}) instead.

The final drop shape is that of a spherical cap. The radius $R^*$ of its
contact line serves as a characteristic length scale. The time
\begin{equation}\tau_0=R^*\xi/ \sigma_{lg}\label{t0}
\end{equation}
defines a characteristic time scale.

When the spherical cap approximation  can be used for the drop shape then
at any moment of time only one parameter is needed to specify
the instantaneous configuration of the drop: either the
time-dependent base radius $R(t)$ or the dynamic contact angle $\theta(t) $.
The drop volume conservation condition implies a
relationship between $R(t)$ and $\theta(t)$:
\begin{equation}
R^3(t) =\frac{3V}\pi \frac{[1+\cos\theta(t)]\sin\theta(t)} {[1-
\cos\theta(t)][2+\cos\theta(t)]} \ .  \label{21}
\end{equation}
Thus Eq.~(\ref{clv}) leads to the following ordinary differential
equation for the dynamic contact angle $\theta ( t)$
\cite{RCVC00}:
\begin{equation}
\frac{d\theta }{dt}=\left( \frac{\pi}{3V}\right)^{1/3} \left\{
[1-\cos\theta(t)][2+\cos\theta(t)]^2\right\} ^{2/3}[\cos\theta(t)-
\cos\theta_{eq}]\ . \label{22}
\end{equation}
Note, that the well known dependencies, $\theta \left( t\right)
\sim t^{-3/7}$ and $R\left( t\right) \sim t^{1/7}$, (see, e.g.,
\cite{RCO99,RCVC00,Rieutord}) are asymptotic solutions of
Eqs.~(\ref{21}, \ref{22}) for small contact angles.

Nikolayev and Beysens \cite{VNDB1} considered the relaxation of an
elongated drop by assuming its surface to be a part of a spheroid
at any time moment. The contact line is then ellipse with
half-axes $R^{*}(1-r_x)$ and $R^{*}( 1+r_y)$ where the relative
deviations $r_x$ and $r_y$ were assumed to be small, $0<r_x,r_y
<<1$. Such an approximation can be adequate at the end of the
relaxation. However, it allowed only the case $\theta _{eq}<\pi
/2$ to be considered. Nikolayev and Beysens obtained exponential
asymptotic solutions for $r_x(t)$ and $r_y(t)$. Two relaxation
times were identified. One of them appears when the drop surface
is a spherical cap, i.e., when $r_x(0)=-r_y^(0)$:
\begin{eqnarray}
\tau _s=\tau _0/\left[ \sin ^2\theta _{eq}\left( 2+
\cos \theta _{eq}\right) \right].\label{25}
\end{eqnarray}
When the initial contact line is an ellipse with $r_x(0)=r_y(0)$,
the relaxation time obtained using spheroidal approximation reads
\begin{eqnarray}
\tau _n=45\tau _0\left( 1+\cos \theta _{eq}\right) /\left[ \left(
108+41\cos \theta _{eq}+14\cos ^2\theta _{eq}+17\cos ^3\theta
_{eq}\right) \left( 1-\cos \theta _{eq}\right)
 \right] .\label{26}
\end{eqnarray}

\section{ Description of the numerical algorithm}

The following numerical algorithm was implemented. First, for a
given position of the contact line and  fixed volume $V$ the
equilibrium drop shape is determined. Then the normal projection
of the velocity at every point of the contact line is obtained by
the help of Eqs.~(\ref{var3p}, \ref{vl2s}).
 Next, from the kinematics condition
\begin{eqnarray}
{{d\vec{R}}\over {dt}} =\vec{v}_n ,
\end{eqnarray}
the contact line position at the next instant of time is
found explicitly. The above algorithm is
repeated for the successive time steps.

The main ingredients of this algorithm are the determination of
the equilibrium drop shape with given volume and given contact
line, and the calculation of the velocity of the contact line. The
drop shape algorithm is essentially an iterative minimization
procedure based on the local variations method \cite{SI1}. Here,
only a very concise description will be given; more details can be
found in \cite {SI2}. The drop shape is approximated by a set of
flat triangles with total of $N=12781$ vertex points, $N_L=360$ of
these are located at the contact line (see Fig.~\ref{fig4}). For a
given contact line, the area of the drop surface is expressed in
terms of the coordinates of the $N$ points. The change of the drop
shape is achieved by approximation of the virtual displacements.
In the $3N-3N_L$ coordinate space, the set of all possible
displacements of $N-N_L$ points is considered while keeping the
volume and the contact line constant. We use the Monte Carlo
scheme for choosing the points which we will try to move. At every
iteration step the drop shape is changed in such a way that the
free energy decreases while the drop volume is kept constant. Thus
eventually the minimal drop surface is found.

The approximation of the normal projection of the velocity of the
contact line at each of the $N_L=360$ vertex points  of the
contact line is obtained by solving the finite approximation of
Eq.~(\ref{var3p}). The method takes into account that the finite
approximation of Eq.~(\ref{var3p}) is described by energy and
volume variations under displacements of these points. The
correctness of the obtained solution at every time step is checked
by keeping track of the accuracy with which the coordinates of the
points from the surface satisfy the Laplace condition and
Eq.~(\ref {clv}). For given contact line and volume, the initial
approximation of the drop shape is found in the following way.
First, for the given volume we find the spherical cap
approximation. Then we perform an iterative procedure  which
transforms the contact line gradually while the volume is kept
fixed until the desired contact line is obtained.

In order to ensure  better work of the minimization procedure, we
perform regular check of the surface mesh and re-adjust the mesh
to keep the approximation of the liquid/gas interface uniform.
This allows us to maintain high accuracy in determining the
contact angle with an error of the order of $0.01^\circ$. At a
given contact line node point the contact angle is defined as the
angle between the plane of the substrate and the plane of the
triangle whose corner coincides with that point.

\section{Results and Discussion}

\subsection{Spherical cap relaxation}

To test the described above 3D code, we check it against the
numerical solution of Eqs.~(\ref{21},\ref{22}) obtained for a
broad interval of values of the equilibrium contact angle
$\theta_{eq}$. The initial contact line radius differs from its
equilibrium (final) value  $R^*$, the deviation being $\Delta
R_0=R(0)-R^*$. As follows from Eq.~(\ref {t0}), we  can set
$R^*=1$ and $\tau _0=1$ without a loss of generality.

\begin{figure}[tb]
\includegraphics[height=7cm]{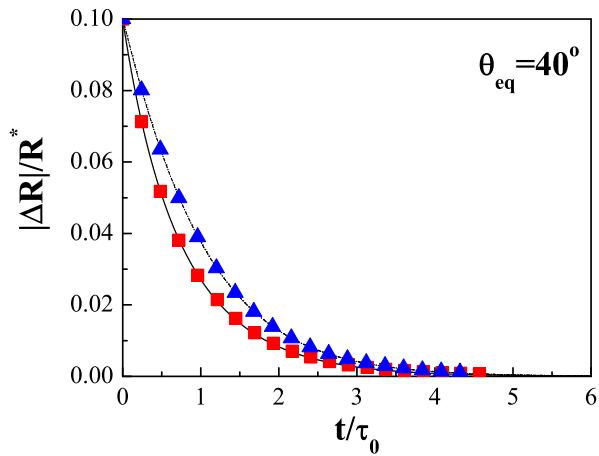}
\caption{\label{fig1}(Color online) Time dependence in $\tau_0$
units of the absolute value of the deviation of the contact line
radius from the equilibrium value $|\Delta R(t)|$ in $R^*$ units
for $\theta_{eq}=40^\circ$ calculated for a drop with a spherical
cap shape. Solid and dashed lines: solutions of Eqs.~(\ref{22}),
(\ref{21}) for $R(0)<R^*$ and $R(0)>R^*$ respectively. Squares and
triangles: numerical 3D calculations for $R(0)<R^*$ and $R(0)>R^*$
respectively (for convenience, every $20^{\textrm{th}}$ data point
is displayed).}
\end{figure}

A comparison of the numerical data, obtained by both methods and
displayed in Fig.~\ref{fig1}, shows a very high (less than 1\%)
accuracy of the 3D code. It can be seen from Fig.~\ref{fig1} that
for the same values of $\theta_{eq}$ and $|\Delta R_0|$  the
solutions for receding contact line, $R(0)>R^*$, and advancing
contact line, $R(0)<R^*$, differ. This follows directly from
Eqs.~(\ref{clv}) and (\ref{21}) since the following inequality
holds
\begin{eqnarray}| \cos \theta ( R^*) -
\cos ( \theta ( R^*) +\delta \theta) | \neq |
\cos \theta ( R^*) -\cos ( \theta ( R^*) -\delta
\theta) | \label{dcos}
\end{eqnarray}
By substituting this inequality in Eq.~(\ref{clv}) it follows that
 for the same absolute value of the deviation $|\Delta R_0|$
there is a difference in the initial velocities for advancing and
receding contact lines.

We studied the possibility to fit the obtained numerical solutions
for $R(t)$ by  power and exponential functions. We use the
following definition of the relative error of the fit
$\overline{R(t)}$ with respect to $R(t)$:
\begin{eqnarray} 
\Delta =\max^{t^*}_{t=0} \left( \frac{\left| R\left( t\right)
-\overline{R\left( t\right) }\right| }{\left| R\left( t\right)
-R^{*}\right| }\right)
\end{eqnarray}
For small initial deviations $|\Delta R_0|$, it turns out that the
exponential fit with
\begin{eqnarray}
\overline{R\left( t\right) }=R^*+|\Delta R_0|\exp(-t/\tau) \ ,
\label{expf}
\end{eqnarray}
\begin{figure}[tb]
\includegraphics[height=7cm]{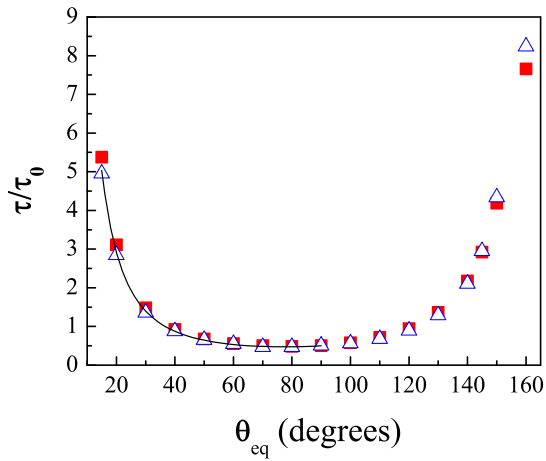}
\caption{\label{fig2}(Color online) The spherical cap relaxation
time $\tau$ in $\tau_0$ units
 as function of the equilibrium contact angle
$\theta_{eq}$ for initial deviation $|\Delta R_0|=0.03$ in $R^*$
units: the solid squares are the results for $\Delta R_0=0.03$;
the empty triangles are the results for $\Delta R_0=-0.03$ and the
solid line is $\tau_s$ (Eq.~(\ref{25})) for
$\theta_{eq}<90^\circ$.}
\end{figure}
where $\tau$ is the only fitting parameter,  describes very well
the data for all studied values of $\theta_{eq}$. The relaxation
time $\tau$ depends on the initial deviation $\Delta R_0$ and when
$|\Delta R_0| \to 0$, $\tau$ tends to the spherical relaxation
time $\tau_s$ (Eq.~\ref{25}).

We first obtained the  solutions for $|\Delta R (t)|$ by the 3D
numerical simulation for initial deviation $|\Delta R_0|=0.03$ and
for contact angles $15^\circ\leq \theta_{eq} \leq 165 ^\circ$. By
fitting the obtained solutions with exponential decay function we
determined the corresponding relaxation times $\tau$ as function
of the equilibrium contact angle $\theta_{eq}$ in the above
interval of values. This dependence is shown in Fig.~\ref{fig2}:
the squares are the results for $\Delta R_0=-0.03$ and the open
triangles are for $\Delta R_0=0.03$. The thin solid line in the
figure is the spherical relaxation time $\tau_s$  (see
Eq.~(\ref{25})) in the interval $\theta _{eq}<90^\circ$. The
exponential approximations of the solutions are obtained in the
time interval $[ 0,t_{end}^{100}] $ determined so that $\left|
R\left( t_{end}^{100}\right) -R^{*}\right| = 0.01\left| R\left(
0\right) -R^{*}\right|$, that is the amplitude of the initial
deviation has decreased hundred times. The exponential
approximation is obtained under the condition that it coincides
with the numerical solution at the initial and final points,
$\{0,t_{end}^{100}\}$. The maximal relative deviation of the
obtained exponential approximations from the numerical solutions
does not exceed $\Delta< 3\%$. When $|\Delta R_0|$ decreases the
precision of the exponential approximation increases. When
$|\Delta R_0|$ increases, e.g. $|\Delta R_0|=0.1,0.2$, the
precision of the exponential approximation to the  numerical
solution of Eqs.~(\ref{22}, \ref{21}) in the time interval $\left[
0,t_{end}^{100}\right] $ decreases.

When the equilibrium contact angle $\theta_{eq}$ increases  the
relative deviation $\Delta$ decreases. The cases of advancing and
receding contact lines differ with less than $1-2\%$ for
$\theta_{eq}\geq 40^\circ$. Also when $|\Delta R_0|$ increases, so
does the deviation of the relaxation exponent $\tau$ (Eq.~\ref
{expf}) from the spherical relaxation time $\tau_s$. When the
exponential approximation in the interval
$\left[0,t_{end}^{100}\right]$ becomes unacceptable, e.g., when
$\Delta$ more than $3\%$, or $-\frac{1}{3}R^*\leq\Delta R \leq
3R^*$ then a good approximation could be obtained either by
splitting the interval $\left[ 0,t_{end}^{100}\right]$ into
several subintervals and approximating the numerical solution on
every such subinterval with an exponential function with a
specific relaxation time $\tau$ or by fitting the numerical
solution with a second or higher order exponential decay function.
For example, for the considered cases $|\Delta R_0|=0.1,0.2$ the
fit with an exponential decay function of the second order
\begin{eqnarray}
\overline{R\left( t\right) }=R^*+a_1\exp(-t/\tau_1)+
a_2\exp(-t/\tau_2)\ ; |a_1|\geq |a_2| \ ,
\label{expf2}
\end{eqnarray}
where $a_1,\tau_1,a_2,\tau_2$ are the fitting parameters, on the interval
$\left[0,t_{end}^{100}\right]$
 becomes much better than with the first order exponential decay function
(Eq.~(\ref{expf})) especially for $\theta_{eq}<40^\circ$. For
example for $\theta_{eq}=40\raisebox{1ex}{\scriptsize o}$ and
$\Delta R_0=-0.1$ the maximal deviation with Eq.~(\ref{expf2}) is
less than $1\%$ as compared to $10\%$  with Eq.~(\ref{expf}).
\begin{table}[htb]
\begin{tabular}{|c|c|c|c|c|c|c|}
\hline
&&&&&&\\[-20pt]
&$\tau_s/\tau_0$ & $a_1$ & $\tau_1/\tau_0$ & $a_2$ & $\tau_2/\tau_0$ & $\Delta$  \\[1pt]
\hline
&&&&&&\\[-20pt]
$\theta_{eq}=10^\circ$  &$11.1$   &$-0.08$    &10.8   &$-0.02$    &3.9
   &2.7\%\\[1pt]
\hline
&&&&&&\\[-20pt]
$\theta_{eq}=40^\circ$  &$0.87$   &$-0.084$   &0.866  &$-0.016$   &0.35
  &1\%\\[1pt]
\hline
&&&&&&\\[-20pt]
$\,\theta_{eq}= 70^\circ$  &\,$0.48$ & -0.092 & 0.484 & -0.008
& 0.248 & 0.08\% \\[1pt]
\hline
\end{tabular}
\caption{Relative deviation $\Delta$ of the exponential
approximation of second order}\label{T2}
\end{table}

As can be seen from Table \ref{T2} , $\tau_1$ is close to $\tau_s$
and the amplitude $a_2$ is sufficiently large so that the
influence of the second exponent should not be neglected. When the
equilibrium contact angle $\theta_{eq} \to \pi/2$ the second
amplitude
 $a_2$ decreases. For contact angles $\theta_{eq} \in (0,\pi/2]$
the amplitude $a_2$ in the case $\Delta R_0=0.1$ is smaller than
in the case $\Delta R_0=-0.1$. For contact angles $\theta_{eq} > \pi/2$
 the opposite is true.

For small contact angles, e.g., $\theta_{eq}=3^\circ, 5^\circ$ we tried
to fit our data also with a power
function $f\sim t^{1/7}$. It appears that it is possible to find
a time interval at the beginning
where the numerical data is well described by the power function
 but the overall behavior is still
better described by the exponential approximation.

\subsection{Relaxation of elongated drops}

\begin{figure}[tb]
\includegraphics[height=7cm]{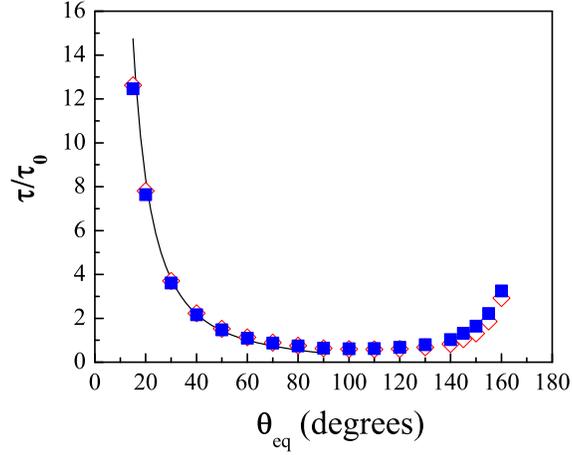}
\caption{\label{fig3}(Color online) The relaxation time for the
elongated drop  in $\tau_0$ units as a function of the equilibrium
contact angle $\theta_{eq}$ for $r_y(0)=r_x(0)=0.03$ in $R^*$
units: the solid squares and empty diamonds are the results for
the exponential fits of $r_x(t)$ and $r_y(t)$ respectively. The
solid line is $\tau_n$ (Eq.~(\ref{26})) for $\theta_{eq}
<90^\circ$.}
\end{figure}

Here we consider the relaxation of a liquid drop when the
initially elliptical contact line (with initial deviations
$r_x(0)=r_y(0)=|\Delta R_0|>0$) relaxes towards circular contact
line. We study the time relaxation $r_x(t)$ and $r_y(t)$ of the
two extreme points $M$ and $N$ of the ellipse, where
$R^{*}(1-r_x)$ and $R^{*}(1+r_y)$ are the half-axes of the contact
line ellipse. The goal is to check the validity of the spheroidal
approximation in \cite{VNDB1} and extend the results to the domain
$\theta >90^\circ$. The analysis of the data obtained by the
method described in Section 3 shows that the time relaxation for
initial deviations up to $r_x(0)=0.2$ is again well described by
an exponential decay function of the first or second order (i.e.
by the sum of two exponential functions with different relaxation
times) in the time interval $[0,t_{end}^{100}]$. The error of the
fit is $\Delta <3\%$. The obtained values for the relaxation time
$\tau$ (Eq.~(\ref{expf})) for contact angles in the interval
$15^\circ\leq \theta \leq 165^\circ$, $r_x(0)=0.03$, are shown in
Fig.~\ref{fig3}. For $15^\circ\leq \theta _{eq}\leq 50^\circ$ the
relative deviation from Eq.~(\ref{26})
is of the order of $2-4\%$. Outside of this interval it increases
fast and for $\theta _{eq}\sim 90^\circ$  it reaches $\sim 60\%$.
The increase of the deviation is due to the fact that the
approximation of the spheroidal cap to the quasi-stationary drop
shape is worsening with the increase of the contact angle $\theta
_{eq}$. Note that while the surface curvature $k$ has to remain
constant along the surface according to Eq.~\ref{Lap}, it varies
as much as 20\% for the spheroid with $r_x(0)=0.1$. In the 3D
simulation, the curvature variation along the surface is less than
0.5\% which is a good accuracy.
\begin{figure}[tb]
\includegraphics[height=7cm]{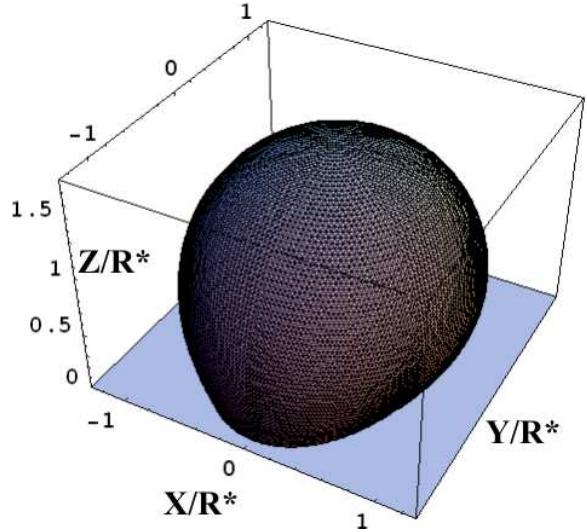} 
\caption{\label{fig4} (Color online)  The initial drop shape with
elliptical contact line and minimal surface for
$\theta_{eq}=120^\circ$, $r_x(0)=0.2$ in $R^*$ units and volume
$V/{R^*}^3=5.44$. }
\end{figure}

\begin{figure}[tb]
\includegraphics[height=7cm]{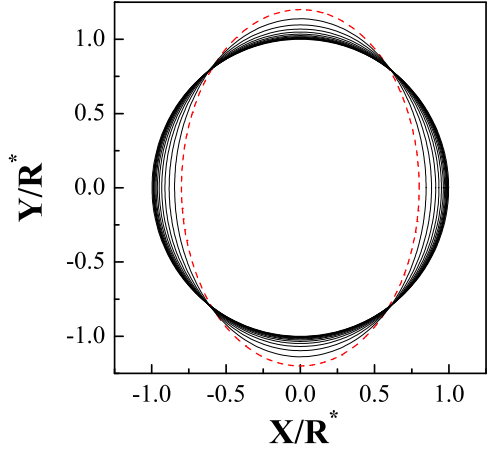}
\caption{\label{fig5}(Color online)  The contact line positions
obtained with time step $(0.2\tau_0)$ for $\theta_{eq}=120^\circ$
and $r_x(0)=0.2$ in $R^*$ units. The dashed line is the initial
position. }
\end{figure}

\begin{figure}[tb]
\includegraphics[height=7cm]{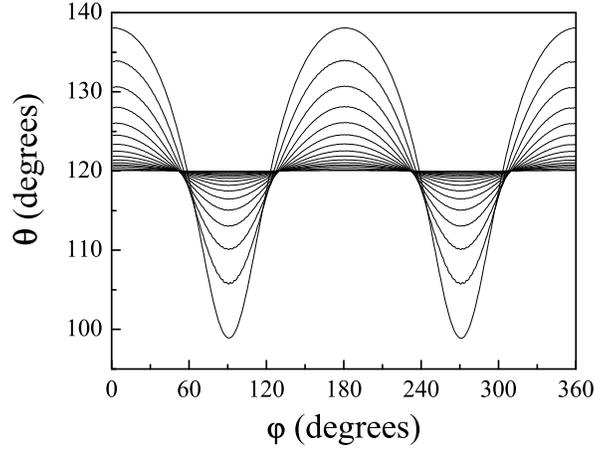}
\caption{\label{fig6}  The contact angle as a function of the
polar angle $\varphi$ at successive moments of time $\left\{
0,0.2n,n=1,2,...\right\} $ in $\tau_0$ units for
$\theta_{eq}=120^\circ$ and $r_x(0)=0.2$.}
\end{figure}

\begin{figure}[tb]
\includegraphics[height=7cm]{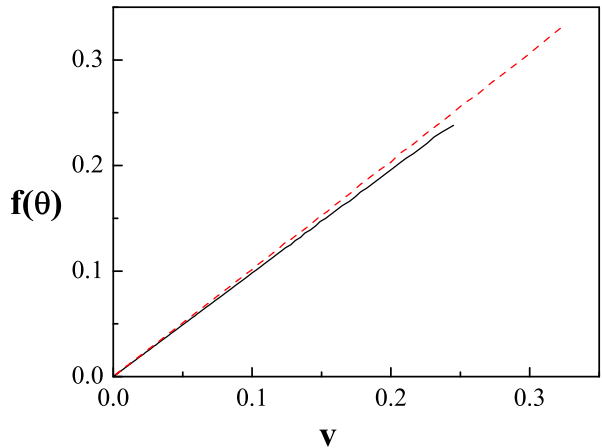}
\caption{\label{fig7}(Color online) The dependence of the function
$f(\theta)=|\cos\theta(t)-\cos\theta_{eq}|$ on the contact line
velocity in two contact line points for $\theta_{eq}=120^\circ$
and $r_x(0)=0.2$ in $R^*$ units. The solid line corresponds to
$v=dr_y/dt$, and dashed line to $v=dr_x/dt$ (in $R^*/\tau_0$
units).}
\end{figure}

The numerical results for $\theta _{eq}= 120^\circ$ and $r_x(0)=0.2$,
are shown in Figs. \ref{fig4}-\ref{fig7}.
The results for other contact angles look qualitatively the same way. The
initial drop shape is shown in Fig.~\ref{fig4}. The volume of the drop is
chosen so that the final shape is the spherical cap with
the radius of the contact line $R^*=1$ and a contact angle
$\theta_{eq}=120^\circ$. The contact line evolution is shown
in Fig.~\ref{fig5}.  The time evolution of the contact angle
along the contact line is shown in Fig.~\ref{fig6}.

The algorithm efficiency can be checked against Eq.~(\ref{clv})
which was not directly used. Fig.~\ref{fig6} shows how good the
algorithm precision is: the difference between the slopes of the
two straight  lines is less than $2\%$.

Note that for equal initial deviations
$r_x(0)=r_y(0)$ at $M$ and $N$
the initial contact
angles and the initial velocities at both points are different.
From the fact that the relaxation times for both $r_x$ and $r_y$
are close (when exponential approximation Eq.~(\ref{expf}) is used)
it does not
follow that the velocities of both points are close as it would seem
if one simply differentiates Eq.~(\ref{expf}) with respect to
time $t$. This can be seen if one examines carefully Figs.
\ref{fig6} and \ref{fig7}.
When the initial deviations are in the interval
$(-\frac{1}{3}R^*, 3R^* )$ then a good approximation
could be obtained either by
splitting the time interval  into several subintervals and
approximating the numerical solution on
every such subinterval with an exponential function with a
specific relaxation time $\tau$ or by
fitting the numerical solution with a sum of two or
more exponential functions.

\subsection{Drops of complicated shapes}

\begin{figure}[tb]
\includegraphics[height=7cm]{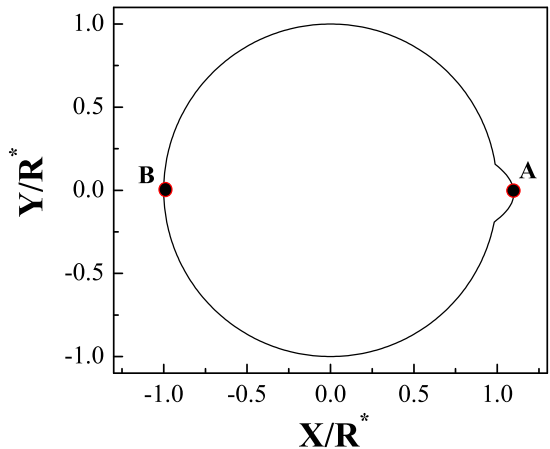}
\caption{\label{fig8}(Color online) The contact line of a drop which is
almost a spherical cap with a small
deformation around one point.}
\end{figure}

\begin{figure}[tb]
\includegraphics[height=7cm]{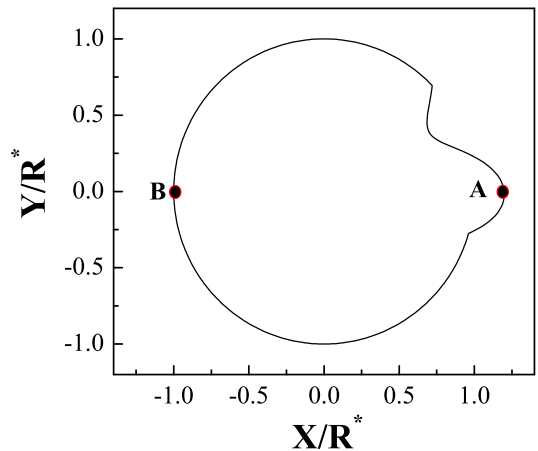}
\caption{\label{fig9}(Color online) The contact line of a drop
which is almost a spherical
cap with larger deformation around the point \textbf{A}.}
\end{figure}

\begin{figure}[tb]
\includegraphics[height=7cm]{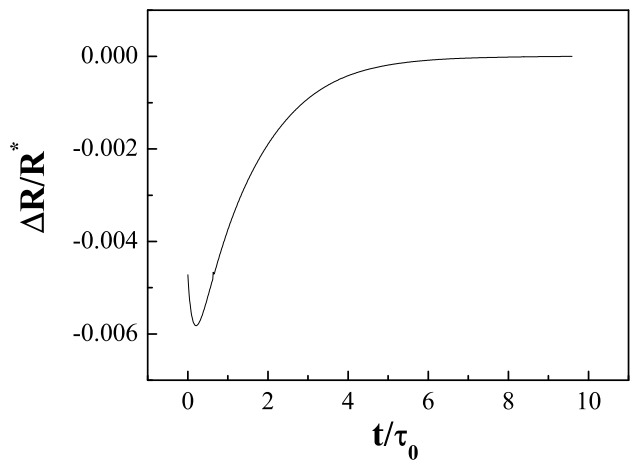}
\caption{\label{fig10} Time dependence in $\tau_0$ units of
$\Delta R(t)$ in $R^*$ units at the point \textbf{B} for a drop
with initial contact line shown in Fig.~\ref{fig9}. }
\end{figure}

We study here the relaxation of drops with some example contact
lines to demonstrate how the relaxation at one point of the
contact line is influenced by the dynamics of the whole contact
line. Consider the relaxation of a drop which is almost a
spherical cap except for a local perturbation around one point of
the contact line. More specifically, let us consider the
relaxation of a drop with a final equilibrium contact angle
$\theta_{eq}=50^\circ$ and with the initial contact line shown in
Fig.~\ref{fig8}.
We find that the time relaxation of the point \textbf{A}$(1.1,0)$
 is well approximated by an exponential
decay function (\ref{expf2}) of the second order: $a_1=0.066$,
$\tau _1=0.163$, $a_2=0.024$, $\tau _2=0.88$ and the relaxation of
the point \textbf{B}$(-1,0)$ by the exponential decay function
(\ref{expf}) of the first order with $\tau=1.05$. All the three
relaxation times $\{0.163,0.88,1.05\}$ differ from each other and
from the relaxation times for spherical and elongated drops
$\tau_s=0.65$, $\tau_n=1.43$ found for $\theta_{eq}=50^\circ$ from
Eqs.~(\ref{25},\ref{26}). It  appears thus that the relaxation of
the point \textbf{B} is influenced by the perturbation around the
point \textbf{A}. Moreover even the type of the relaxation of the
point \textbf{B}, whose neighborhood is a part of circle, is not
universal and depends on the deformation around the point
\textbf{A}. For example when the contact line is of the type shown
in Fig.~\ref{fig9} we obtain that the relaxation of the point
\textbf{B }is as shown in Fig.~\ref{fig10}. It is possible even to
find a deformation around \textbf{A} such that the relaxation of
the point \textbf{B} is practically linear in a broad time
interval.

\section{Conclusions}

We have described a method and applied it  to simulate the
quasi-static relaxation of drops with different initial 3D shapes
starting directly from the variational
principle of Hamilton, taking into
account only the large dissipation in the vicinity of
the contact line during the contact line motion.

We have shown rigorously for arbitrary contact line shape
using the standard mechanical description of dissipative
system dynamics that the  introduction of a  friction dissipation
term proportional to the contact line length
in the case of quasi-static relaxation leads  to the
well known local relation between the contact line velocity and
the dynamic contact angle.

We find in the case of spherical cap approximation that
 the time relaxation of the contact line
radius is very well described by an exponential decay function of
 the first or the second order depending on the magnitude of
the initial deviation. The relaxation time
 $\tau$ depends on the initial deviation $\Delta R_0$ and when
$|\Delta R_0| \to 0$, $\tau$ tends to the spherical relaxation time
$\tau_s$ defined in Ref. \cite{VNDB1}. For higher values of
$|\Delta R_0|$, e.g. $|\Delta R_0|=0.1,0.2$, the data is better
described by the sum of two exponentials with different
relaxation times. The power function fits do not describe well the data.

In the case of elongated drops, the relaxation is again very well
described by an exponential decay function. The
relaxation time is within  2-4\% from that obtained with
the spheroid approximation for the drop shape
\cite{VNDB1} in the range
$15^\circ\leq \theta _{eq}\leq 50^\circ$. For the larger angles,
the relaxation time can only be obtained by the described 3D
numerical simulation.

Previously exponential relaxation is found in some
experimental studies, e.g., in \cite{N68} and more recently in
\cite{JFM02}. Theoretically, exponential relaxation is found in
\cite{VNDB1} and asymptotically at  long times in \cite{RCO99},
 as well as in the Monte Carlo simulations of
the Ising model for drop spreading \cite{NJ04}.

By simulating the relaxation of drops of complicated 3D shape, we
showed that, although the local Eq.~(\ref{clv}) is satisfied, the
relaxation at a given point of the contact line is influenced by
the relaxation dynamics of the whole drop surface. This is a
manifestation of the non-local character of the contact line
motion.

\bibliography{dynamics6V}
\end{document}